\documentclass[preprint,aps,prd,showpacs,superscriptaddress,nofootinbib]{revtex4}

\usepackage{amsfonts}
\usepackage{multirow}
\usepackage{mathrsfs}
\usepackage{graphicx}
\usepackage{amsmath}
\usepackage{amssymb}
\usepackage{bm}
\usepackage{bbm}
\usepackage{slashbox}

\def\OMIT#1{}

\newcommand{\nn}{\nonumber}

\newcommand{\beq}{\begin{equation}}
\newcommand{\eeq}{\end{equation}}
\newcommand{\bqa}{\begin{eqnarray}}
\newcommand{\eqa}{\end{eqnarray}}

\begin{document}
\title{\mbox{}\\[10pt]
Short-range force between two Higgs bosons}


\author{Feng Feng\footnote{Current address: School of Science, China University of Mining and Technology, Beijing 100083, China. Email: fengf@ihep.ac.cn}}
\affiliation{School of Physical Science and Technology, Southwest
University, Chongqing 400700, China\vspace{0.2cm}}
\affiliation{Center for High Energy Physics, Peking University,
Beijing 100871, China\vspace{0.2cm}}
\author{Yu Jia\footnote{Email: jiay@ihep.ac.cn}}
\affiliation{Institute of High Energy Physics, Chinese Academy of
Sciences, Beijing 100049, China\vspace{0.2cm}}
\affiliation{Theoretical Physics Center for
Science Facilities, Chinese Academy of Sciences, Beijing 100049,
China\vspace{0.2cm}}
\affiliation{Center
for High Energy Physics, Peking University, Beijing 100871,
China\vspace{0.2cm}}

\author{Wen-Long Sang \footnote{Email: wlsang@ihep.ac.cn}}
\affiliation{School of Physical Science and Technology, Southwest
University, Chongqing 400700, China\vspace{0.2cm}}
\affiliation{State Key Laboratory of Theoretical Physics, Institute of Theoretical Physics, Chinese Academy of Sciences,
Beijing 100190, China\vspace{0.2cm}}

\date{\today}
\begin{abstract}
The $S$-wave scattering length and the effective range of the Higgs
boson in Standard Model are studied using effective-field-theory
approach. After incorporating the first-order electroweak
correction, the short-range force between two Higgs bosons remains
weakly attractive for $M_H=126$ GeV. It is interesting to find that
the force range is about two order-of-magnitude larger than the
Compton wavelength of the Higgs boson, almost comparable with the
typical length scale of the strong interaction.
\end{abstract}




\maketitle

\noindent{\it Introduction.} The ground-breaking discovery of a new
particle with mass around 126 GeV by the \textsc{Atlas} and
\textsc{CMS} Collaborations at CERN Large Hadron Collider (LHC) in
July 2012~\cite{Aad:2012tfa,Chatrchyan:2012ufa}, heralds an exciting
new era of particle physics. Undoubtedly, the top priority in the
coming years is to pin down the detailed property of this new boson
as precisely as possible, {\it e.g.}, its quantum number, decay and
production patterns~\cite{Dawson:2013bba}. Hopefully, one will
finally be able to determine whether this new boson is the
long-sought Brout-Englert-Higgs boson of
Standard Model (SM)~\footnote{For simplicity, hereafter we shall simply call it Higgs boson.} or of some exotic origin.

The SM Higgs boson is an elementary scalar particle carrying
$J^{PC}=0^{++}$. An enormous amount of work has been devoted to
exploring the physics involving an individual Higgs boson, while the
respective studies concerning the multi-Higgs-boson dynamics, such
as double- or triple-Higgs productions at LHC experiments, are still
in the infancy stage~\cite{Dawson:2013bba}. Nevertheless, a thorough
investigation of the latter is crucial in unraveling the nature of
the Higgs potential since it directly probes the self-coupling of
the Higgs bosons.

It is of fundamental curiosity to inquire the short-range force
which two Higgs bosons would experience. A few decades ago, Cahn and
Suzuki~\cite{Cahn:1983vi}, as well as Rupp ~\cite{Rupp:1991bb},
studied the interaction between two Higgs bosons by utilizing some
nonperturbative methods, only including the Higgs self coupling.
They claimed that the attraction would become strong enough as
$M_H>1.3$ TeV to bind them together into a Higgs-Higgs bound
state (Higgsium), albeit highly unstable. Such a large Higgs mass
violates the perturbative unitarity bound~\cite{Lee:1977yc}. If the
new particle discovered in LHC is indeed the SM Higgs boson, the
Higgsium seems unlikely to be formed in the first place. This
expectation is supported by the recent lattice simulation of the
electroweak gauge model~\cite{Wurtz:2013ova}. Nevertheless,
Grinstein and Trott recently suggested that the possibility for the
existence of the light Higgsium might be still open due to some new
physics scenario at TeV scale~\cite{Grinstein:2007iv}.

The model-independent parameters that characterize any short-range
force are {\it scattering length} and {\it effective range}. The
effective-field-theory (EFT) approach provides a systematic
framework to expedite inferring these parameters. The aim of this
paper is to decipher the short-range force experienced by two God
particles following this modern doctrine. In particular, we will
investigate the influence of the $W$, $Z$ and top quark on the
inter-Higgs force. It will be interesting if our predictions can be
confronted by the future lattice simulation, or even by the double
Higgs production experiments.

\vspace{.2 cm} \noindent{\it The Higgs sector in SM.} After the
spontaneous electroweak symmetry breaking, the Higgs sector in SM
Lagrangian reads (in unitary gauge):
\bqa
& & {\cal L}_{H} = {1\over 2}(\partial_\mu H)^2- {1\over 2} M_H^2
H^2 -{\lambda v\over 4}H^3-{\lambda\over 16} H^4
+ {2M_W^2\over v} W^{+\mu}W^-_{\mu}H + {M_W^2\over v^2}
W^{+\mu}W^-_\mu H^2
\nn\\
&& + {M_Z^2\over v} Z^\mu Z_\mu H + {M_Z^2\over 2 v^2} Z^\mu Z_\mu
H^2 -{m_t\over v} \bar{t}t H+\cdots,
\label{Lagrangian:H:Sector}
\eqa
where $v\approx 246$ GeV is the vacuum expectation value of the
Higgs field, $M_H$, $M_W$, $M_Z$, $m_t$ signify the masses of the
Higgs boson, $W^\pm$, $Z$, and the top quark, respectively. All the
other fermions are neglected due to much weaker Yukawa coupling. We
follow the convention of parameterizing the Higgs potential as in
Refs.~\cite{Hollik:1988ii,Denner:1991kt} such that
$M_H=\sqrt{\lambda\over 2} v$. For a light 126 GeV Higgs boson, the
self coupling $\lambda\approx 0.52$, and we have an entirely
weakly-coupled Higgs sector.

\vspace{.2 cm}\noindent{\it Nonrelativistic EFT for Higgs boson.} We
are interested in the near-threshold elastic scattering between two
Higgs bosons, thereby only the $S$-wave channel needs be retained.
Since the momentum of each Higgs boson is much lower than the
remaining mass scales $M_H\sim M_W\sim M_Z\sim m_t\sim v$, it seems
legitimate to integrate out the contribution from all the
relativistic (hard) modes, and construct the following low-energy
EFT which only involves the nonrelativized Higgs
field~\cite{Jia:2004qt}:
\bqa
&& {\cal L}_{\rm NREFT} = \Psi^* \bigg( i\partial_t + {\nabla^2\over
2 M_H}-
{\partial_t^2\over 2 M_H}\bigg) \Psi  -{C_0\over 4} (\Psi^* \Psi)^2
  -{C_2\over 8} \nabla(\Psi^* \Psi)\cdot \nabla(\Psi^* \Psi)+ \cdots,
\nn\\ \label{NR:EFT:lagrangian} \eqa
where $\Psi^{(*)}$ field annihilates (creates) a Higgs boson. The
126 GeV Higgs boson appears to have a narrow width ($< 10$
MeV)~\cite{Barger:2012hv} so that we treat it as a stable particle.
This EFT is organized by a velocity expansion, and remains valid as
long as the momentum carried by the Higgs boson is smaller than the
UV cutoff of this NREFT, $\Lambda$, which {\it a priori} is expected to be
of the same size as the inverse of the force range, $\sim 1/r< M_H$. The
$S$-wave scattering is mediated by the two 4-boson operators with
the Wilson coefficients $C_0$ and $C_2$. By naturalness one assumes
$C_0\sim {4\pi\over M_H\Lambda}$, $C_2\sim {4\pi\over
M_H\Lambda^3}$. The two-body operator containing $\partial_t^2$
in the parenthesis of (\ref{NR:EFT:lagrangian}) signals the relativistic correction.
With this specific form, the Higgs state in our NREFT is understood
to tacitly obey the relativistic normalization condition, {\it
i.e.}, $\langle H({\bf k}) |H({\bf p})\rangle= \sqrt{{\bf
k}^2+M_H^2}/M_H (2\pi)^3 \delta^3({\bf p}-{\bf k})$.

It is worth emphasizing the legitimacy of integrating out $W$, $Z$
and $t$ in our NREFT. Suppose in a fictitious world in which
$M_W(M_Z,m_t)$ were very close to $M_H$, or $M_W(M_Z,m_t)$ were
roughly half of $M_H$, one would be forced to retain the
nonrelativistic $W(Z,t)$ fields as active degree of freedom in
(\ref{NR:EFT:lagrangian}), in order to properly account for the
near-threshold reactions such as $H\to WW(ZZ,t\bar{t})$, $HH\to
WW(ZZ,t\bar{t})$. Fortunately, in the real world, none of the above
coincidence arises, so we are justified to only keep the
nonrelativistic Higgs field in the NREFT.

The $S$-wave amplitude can be calculated order by order in velocity
(loop) expansion from (\ref{NR:EFT:lagrangian}), with the UV
divergence conveniently subtracted in the MS scheme. In a NREFT that
only contains contact interaction, an all-order result is available
by summing the infinite number of bubble diagrams as a geometric
series~\cite{Kaplan:1998tg,vanKolck:1998bw}. Remarkably, that
nonperturbative result is only subject to slight change when the
relativistic correction is included~\cite{Jia:2004qt}:
\bqa
{\cal A}^{\rm S-wave}_{\rm NREFT} &=& -\bigg[{1\over C_0+ C_2
k^2+\cdots}+ {iM_H\over 8\pi}\gamma^{-1}k \bigg]^{-1},
\label{S:wave:ampl:nonperturbative}
\eqa
where $k$ denotes the momentum in the center-of-mass frame, and
$\gamma \equiv \sqrt{1+k^2/M_H^2}$ is a Lorentz dilation factor,
which embodies the full relativistic correction.

A traditional way of parameterizing  the $S$-wave elastic amplitude
mediated by a short-range interaction is through the $S$-wave phase
shift:
\bqa
{\cal A}^{\rm S-wave}_{\rm ERE} &=& {8\pi\over M_H}\bigg[k\cot
\delta_0 -i\gamma^{-1} k \bigg]^{-1}= {8\pi\over M_H}\bigg[-{1\over
a_0}+{r_0\over 2} k^2+ \cdots -i\gamma^{-1} k \bigg]^{-1},
\label{ERE:S:wave:modified}\eqa
$\delta_0$ implies the $S$-wave phase shift, while $a_0$ and $r_0$
signify the {\it $S$-wave scattering length} and {\it effective
range}, each of which is physical observable. The second line
specifies the so-called effective range expansion, valid only for
small $k$. Again, a factor of $\gamma^{-1}$ is included to recover
the Lorentz invariance~\cite{Jia:2004qt}.

Equating (\ref{S:wave:ampl:nonperturbative}) and
(\ref{ERE:S:wave:modified}), one determines $a_0$ and $r_0$ via
\bqa
a_0 &=&  {M_H\over 8\pi} C_0,\qquad
r_0 = {16 \pi\over M_H} {C_2\over C_0^2}.
\label{def:a0:r0:S:wave}
\eqa

Our central goal is then to deduce the coefficients $C_0$ and $C_2$
to next-to-leading order (NLO) in electroweak couplings. This can be
achieved through matching the $S$-wave amplitude of $HH\to HH$ in SM
onto that in NREFT to one-loop order.

\begin{figure}[t]
\begin{center}
\includegraphics[scale=0.8]{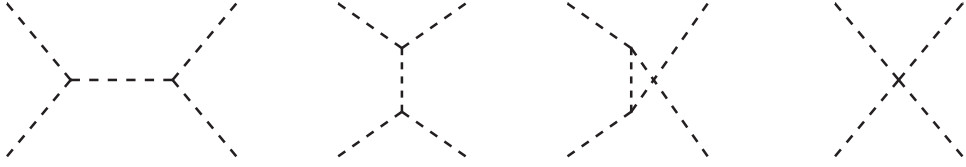}
\caption{The tree-level diagrams for $HH\to HH$ in SM.
\label{feyn:diag:elastic:scattering}}
\end{center}
\end{figure}

\vspace{0.2 cm} \noindent{\it LO results for $a_0$ and $r_0$.} At
tree level, there arise only 4 tree diagrams for the Higgs-Higgs
elastic scattering (in any gauge), as shown in
Fig.~\ref{feyn:diag:elastic:scattering}. Only the physical Higgs
field is involved. The corresponding amplitude is~\cite{Lee:1977yc}
\bqa
&& {\cal A}_{\rm SM}^{(0)} =  - {3M_H^2\over v^2}
\bigg(1+{3M_H^2\over s-M_H^2}+ {3M_H^2\over t-M_H^2}+{3M_H^2\over
u-M_H^2}\bigg)
\nn\\
& & \approx
  {12 M_H^2 \over v^2} \bigg( 1-{8\over 3}
  {k^2\over M_H^2} \bigg)
 + O(k^4),
\label{LO:SM:elastic:amplitude}
\eqa
where $s$, $t$, $u$ are Mandelstam variables. In the second line, we
have carried out the threshold expansion by treating $k^2/M_H^2$,
$t/M_H^2$, $u/M_H^2$ as small perturbations. Near the threshold, the
above expansion automatically projects out the $S$-wave
contribution, up to $O(k^4)$.

The tree-level $S$-wave amplitude in the NREFT side can be obtained
by expanding (\ref{S:wave:ampl:nonperturbative}) accordingly: ${\cal
A}^{(0)}_{\rm S-wave,\:NREFT}=-C_0- C_2 k^2+\cdots$. Comparing it
with (\ref{LO:SM:elastic:amplitude}), one extracts the Wilson
coefficients at LO: $C_0^{(0)} = - {3\over v^2}$, $C_2^{(0)} = {8
\over M_H^2 v^2}$.
Following (\ref{def:a0:r0:S:wave}), we then determine the LO
$S$-wave scattering length the and effective range as
\begin{subequations}
\bqa
a_0^{(0)} &=&  - {3\over 8\pi}{M_H\over v^2}= - {3\over
16\pi}{\lambda \over M_H},
\\
r_0^{(0)} &=&  {128\pi \over 9}{v^2\over M_H^3}= {256\pi\over
9}\left({1\over \lambda}\right){1\over M_H}.
\eqa \label{LO:a0:r0:expressions}
\end{subequations}
We observe that the short-range inter-Higgs force is {\it weakly
attractive}. The magnitude of the scattering length is much smaller
than the Compton wavelength of Higgs boson, while the effective
range is much larger, and ${|a_0|\over r_0}={27\over
1024\pi^2}{M_H^4\over v^4}={27\lambda^2\over 4096\pi^2}\ll 1$.
It is somewhat surprising that the effective range is considerably
($\approx 170$ times) larger than the Compton wavelength of the
Higgs boson, the typical force range of weak interaction. Note that this
situation is drastically opposite to that for the nuclear force,
where the shallow (virtual) bound state arise in the $pn({}^3S_1)$
and $pn({}^1S_0)$ channels due to $|a_0|\gg
r_0$~\cite{Kaplan:1998tg,vanKolck:1998bw}.

For such an unnatural case, we might infer the valid range of our NREFT by
enforcing the convergence of the effective range expansion:
\bqa
{1\over k}\tan \delta_0 = -a_0 \left(1+{a_0 r_0\over 2} k^2+O(k^4)\right).
\eqa
Since ${1\over 2}|a_0| r_0={8\over 3 M_H^2}$, the NREFT seems to apply provided that
$k$ is smaller than the UV cutoff $\Lambda=\sqrt{3\over 8} M_H\approx 77$ GeV. This cutoff value is considerably greater than the naive expectation
of $\Lambda \sim 1/r_0 \approx 1$ GeV.
As a consequence of the wide valid range of this NREFT, it appears to be a virtue to explicitly retain the effect of
relativistic correction as in (\ref{S:wave:ampl:nonperturbative}).

It is instructive to contrast the Higgs model with the simplistic
$\lambda \phi^4$ theory containing a scalar field with mass $m$.
There the inter-particle short-range force is of course repulsive,
and the effective range is about the same order as the Compton
wavelength, quantitatively, $a_0 = {3\over 16\pi}{\lambda \over m}$,
$r_0 = {16\over 3\pi m}+O(\lambda)$~\cite{Jia:2004qt}. We thus infer
that, in the Higgs model, it is the triple Higgs interaction in
(\ref{Lagrangian:H:Sector}) that yields the attractive force and
ultimately wins the competition against the repulsive quartic
interaction. It is also the nontrivial pattern of spontaneous
symmetry breaking that generates the unnaturally large force range.

\vspace{0.2 cm} \noindent{\it NLO results for $a_0$ and $r_0$.} We
wish to assess the impact of the $W$, $Z$ and top quark on the
inter-Higgs force. It is then necessary to incorporate the
first-order electroweak correction to $HH\to HH$. Because the
intermediate $WW,ZZ$ states are permissible to go on-shell, the
coefficients $C_0^{(1)}$ and $C_2^{(1)}$ would become complex, so
are $a_0$ and $r_0$. This situation is analogous to the
nucleon-antinucleon system which can annihilate into multiple
pions~\cite{Haidenbauer:2006dm}.

\begin{figure}[tbH]
\begin{center}
\includegraphics[scale=0.9]{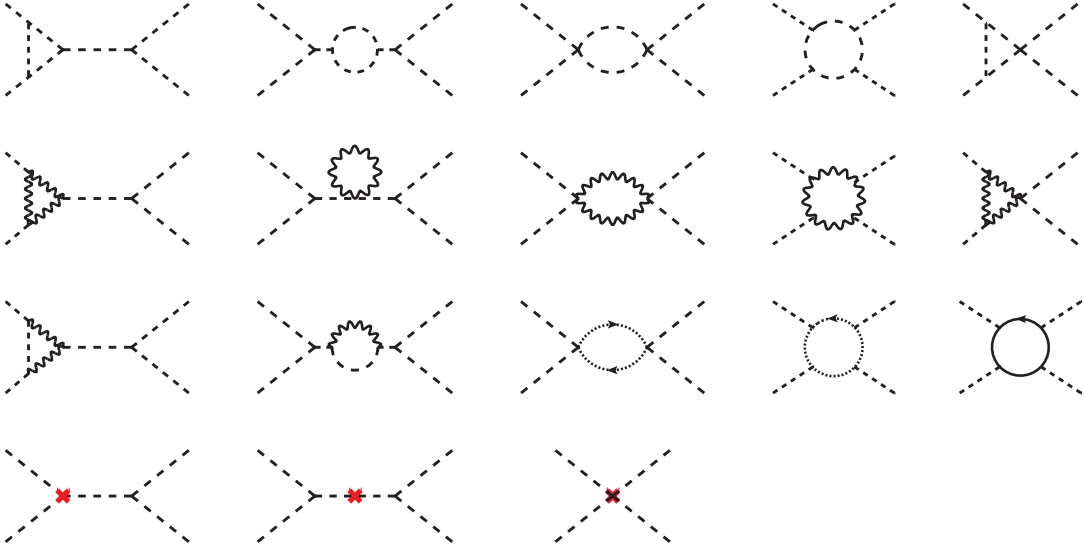}
\caption{Some sample NLO diagrams for $HH\to HH$ in Feynman gauge.
The dashed line stands for the Higgs boson (or the Goldstone bosons
inside the loop), the wavy lines for the $W/Z$ bosons, the dotted
curve for the ghosts, and the solid line for the $t$ quark. The
crosses represent the counterterms for the $H^3$, $H^2$, and $H^4$
vertices, respectively. \label{NLO:diagrams:HH:scattering}}
\end{center}
\end{figure}

We first look at the NREFT side. Expanding the nonperturbative
expression in (\ref{S:wave:ampl:nonperturbative}) to one-loop order,
one finds the $S$-wave amplitude now becomes
\bqa
& & {\cal A}^{(1)}_{\rm S-wave,\:NREFT} = -C_0- C_2 k^2 + i {M_H
k\over 8\pi} \left[ C_0^2 + 2 C_0\left(C_2 - {C_0\over 4
M_H^2}\right)k^2+\cdots \right]+\cdots.
\label{NREFT:amplitude:one:loop} \eqa
The last term stems from the one-loop integration, with the
first-order relativistic correction incorporated. It is odd in
powers of $k$, which is characteristic of the nonrelativistic loop
integration.

We then proceed to compute the first-order electroweak correction to
the near-threshold scattering between two Higgs bosons in the SM
side. There have existed some NLO calculations for $HH\to HH$ with
arbitrary Higgs momentum. However, the results appear to be either
incomplete~\cite{Dvoeglazov:1990bt} or approached in an unrealistic
limit $M_W,M_Z\to 0$~\cite{Gupta:1993tq}.

We choose to work in the Feynman gauge, at a cost of including many
diagrams containing unphysical particles such as the Goldstone
bosons and ghost particles. We use the \textsc{Mathematica} package
\textsc{FeynArts}~\cite{Kublbeck:1990xc} to generate all the Feynman
diagrams and the corresponding amplitudes. For clarity, some sample
diagrams out of the total 603 NLO diagrams are illustrated in
Fig.~\ref{NLO:diagrams:HH:scattering}. We use dimensional
regularization to regularize UV divergence. The package
\textsc{FeynCalc}~\cite{Mertig:1990an} is employed to perform the
tensor reduction.

We choose the standard on-shell renormalization
scheme~\cite{Sirlin:1980nh,Hollik:1988ii,Denner:1991kt} to fix the
counterterms for quadratic, triple, quartic Higgs boson vertices.
Apart from the apparent Higgs wavefunction renormalization constant
$\delta Z_H$ and mass counterterm $\delta M_H^2$, we still need 4
additional renormalization constants, $\delta t$, $\delta s_W$,
$\delta M_W^2$, $\delta Z_e$, representing the counterterms for
Higgs tadpole, Weinberg angle ($s_W\equiv\sin\theta_W$,
$c_W\equiv\cos\theta_W$), $W$ boson mass, and the electric charge,
respectively~\cite{Denner:1991kt}. Some of their analytic
expressions are rather cumbersome.

Fortunately, it is the following specific combination of the
renormalization constants that always enters the expressions for the
$H^3$ and $H^4$ counterterms, which can be recast as
\bqa
\delta Z_e - {\delta s_W \over s_W}- {1\over 2}{\delta M_W^2\over
M_W^2} &= & {\Delta r+\Delta \rho \over 2}-{\alpha\over 8\pi
s_W^2}\bigg(6+{7-4 s_W^2\over 2s_W^2}\ln c_W^2\bigg)
\nn\\
&&  - {1\over 2}{\Sigma_T^{ZZ}(0) \over M_Z^2} - {c_W\over s_W}
{\Sigma_T^{AZ}(0) \over M_Z^2},
\label{final:simplest:combi:counterterm}
\eqa
where $\alpha$ is the fine structure constant, $\Sigma_T^{ZZ}(0)$
and $\Sigma_T^{AZ}(0)$ are the $Z$ boson self energy and the
photon-$Z$ two-point function evaluated at zero momentum, which are
much simpler than $\delta M_W^2$ and $\delta
s_W$~\cite{Denner:1991kt}. $\Delta r$ and $\Delta \rho$, as
constructed out of the meticulous combination of the various gauge
boson self energies, are some familiar UV-finite parameters that can
be directly fixed from the
data~\cite{Sirlin:1980nh,Hollik:1988ii,Denner:1991kt}.

The analytic NLO expression for $HH\to HH$ would be extremely
involved for general kinematics. Fortunately, we are only interested
in its near-threshold behavior. For most diagrams, particularly with
$W$, $Z$, $t$ circulating in the loop, one can simply expand the
integrand in powers of the external momentum $k$, prior to carrying
out the loop integration. This leads to great simplification, since
all the encountered loop integrals then reduce into a set of 2-point
(or less) scalar integrals.

The $s$-channel loop diagrams composed entirely of the Higgs field,
{\it e.g.}, the ones in the first row of
Fig.~\ref{NLO:diagrams:HH:scattering}, deserve some special
attention. Unlike all other diagrams solely dictated by the {\it
hard} region, the nonrelativistic Higgs fields can propagate almost
on-shell in the loop, {\it i.e.}, they also receive the contribution
from the {\it potential} region, which should be fully mimicked by the one-loop diagrams from the
low-energy NREFT.
For these diagrams, we employ the
method of region~\cite{Beneke:1997zp} to extract the contributions
from the hard and potential regions separately. The resulting master
integrals are also the simple 2-point scalar integrals.

Upon summing all the expanded one-loop diagrams and the counterterm
diagrams, the UV divergences are canceled as expected, and one
automatically projects out the $S$-wave contribution. Comparing with
(\ref{NREFT:amplitude:one:loop}), we find its last term is fully
reproduced by the contribution from the potential regions of the
aforementioned $s$-channel diagrams. This can be viewed as a
nontrivial check of our calculation. It is then straightforward to
deduce the NLO coefficients $C_0^{(1)}$ and $C_2^{(1)}$,
subsequently convert into $a_0^{(1)}$ and $r_0^{(1)}$ in line with
(\ref{def:a0:r0:S:wave}), {\it i.e.}, $a_0^{(1)}/a_0^{(0)}=C_0^{(1)}/C_0^{(0)}$, and $r_0^{(1)}/r_0^{(0)}=C_2^{(1)}/C_2^{(0)}-2C_0^{(1)}/C_0^{(0)}$. Conceivably,
their analytic expressions are too lengthy to be reproduced here.

For $M_H=126$ GeV, Eq.~(\ref{LO:a0:r0:expressions}) then implies
that the LO scattering length and effective range are
$a_0^{(0)}=-4.90\times 10^{-5} $ fm, $r_0^{(0)}= 0.267$ fm.

To estimate the NLO contribution, we choose $\alpha=1/137.036$,
$G_F=1.166\times 10^{-5}\;{\rm GeV}^{-2}$, $M_W=80.39$ GeV,
$M_Z=91.188$ GeV, $m_t=173.1$ GeV, $\Delta r=0.0357$, $\Delta
\rho\approx {3 G_F m_t^2\over 8\pi^2\sqrt{2}}=
0.0094$~\cite{Beringer:1900zz}. And the NLO corrections turn out to be
\begin{subequations}
\bqa
a_0^{(1)}/a_0^{(0)} &=&  -0.0355+0.0063 i,
\\
r_0^{(1)}/r_0^{(0)} &=&  0.0245 -0.0145 i.
\eqa
\end{subequations}
The electroweak radiative correction has a modest effect, only
modifying the tree-level result by a few percent. However, the
imaginary parts for both $a_0$ and $r_0$ arise due to the opening of the inelastic channels.
Incorporating the NLO correction, we then predict $a_0= (-4.73\times 10^{-5}-
3.10\times 10^{-7}i)$ fm, and $r_0 =(0.273-3.87\times 10^{-3} i)$ fm.
These are by far the most precise predictions for the
basic parameters that characterize the inter-Higgs force.

To extract the $a_0$ and $r_0$, one needs a very accurate knowledge of the $S$-wave phase shift $\delta_0$ for Higgs-Higgs elastic scattering near threshold. The rather weak inter-Higgs force implies a nearly vanishing $\delta_0$ over a large
momentum range. To determine a (tiny) phase of the $S$-matrix is always a very challenging task. For instance, notwithstanding tremendous efforts, it takes several decades to unambiguously pin down the $\pi\!-\!\pi$
$S$-wave scattering lengths via the interference method~\cite{Gasser:2009zz}.

The double Higgs boson production is one of the important missions
in the next phase of LHC experiment, whose dominant production mechanism is through the parton process
$gg\to H H$ and $W^+W^-/ZZ\to HH$. Due to the low production rates, it perhaps needs a decade to finally observe the double Higgs signals. Even though we could accumulate sufficient amount of double Higgs events near threshold, it would still be difficult to envisage how to extract the phase of $S$-wave scattering amplitude via its interference with the $D$-wave amplitude from this {\it inclusive} production process, not mention the intrinsic uncertainty of parton distribution functions of gluons and $W/Z$ inside the proton. On the other hand, the prospective high-luminosity electron-positron collider may offer much cleaner place to measure the Higgs-Higgs scattering information via the {\it exclusive} processes $e^+e^-\to ZHH,\nu\bar{\nu}HH$, provided that the
sufficient statistics could be achieved.


\vspace{0.2 cm} \noindent{\it Inter-Higgs force in the large
$M_W$,$M_Z$,$m_t$ limit.} It is curious to assess the influence of
$W$, $Z$ and $t$ on the profile of the Higgs-Higgs interaction.
Taking the limit $M_W (M_Z)\to \infty$ (while keep $v$ and $s_W$
intact) and $m_t\to \infty$~\footnote{These limits correspond to
setting the couplings $g_1$, $g_2$, $y_t \longrightarrow\infty$, in
which situation the EW theory ceases to be perturbative.
Nevertheless, our goal is to assess the leading behavior with the
large gauge boson and top quark masses, so we are not too rigorous
here.}, we find asymptotically,
\begin{subequations}
\bqa
&& a_0^{(1)} \to -\frac{9 (2
   M_W^4+M_Z^4)}{64
   \pi ^3 M_H v^4
   } +\frac{9 m_t^4}{16 \pi ^3 M_H v^4},
\\
& & r_0^{(1)} \to  -\frac{32 (2
   M_W^4+M_Z^4)}{9 \pi
   M_H^5
   } +\frac{128 m_t^4}{9 \pi M_H^5
   }.
\label{large:MZ:MTOP:limit}
\eqa
\end{subequations}
where the subleading terms of order $M_Z^2$ and $m_t^2$ are
neglected. The fourth-power mass dependence indicates that the NLO
corrections would rapidly dominate over the LO results as the gauge
boson masses or top quark masses keep increasing, which defies the
decoupling theorem. As $M_W$ ($M_Z$) grow, both $a_0$ and $r_0$
decrease. On the contrary, both $a_0$ and $r_0$ increase with
increasing $m_t$. When $m_t$ crosses around 300 GeV, $a_0$ would
even reverse the sign, so the Higgs-Higgs force would become even
repulsive.

\vspace{0.2 cm} \noindent{\it Summary.} For the first time, we have
thoroughly investigated the profile of the short-range force between
two SM Higgs bosons within the modern EFT context, deducing the
$S$-wave scattering length and the effective range by including the
first-order electroweak correction. The impact of $W$, $Z$ and $t$
on these parameters is addressed. The inter-Higgs force is extremely
weak, yet attractive. But the force range is as large as 0.3 fermi,
comparable with the typical range of the QCD force. It will be
interesting, albeit challenging, if the future lattice simulation
can test our predictions. It might also be of phenomenological
incentive to transplant our analysis to some classes of new physics models.

\begin{acknowledgments}
We thank E.~Radescu for participating in the initial stage of this
work. The research was supported in part by the National Natural
Science Foundation of China under Grant No.~10935012 and No.~11347164, DFG and NSFC
(CRC 110). The research of W.~-L.~S was also supported by the
Fundamental Research Funds for the Central Universities under Grant No.~ SWU114003.
\end{acknowledgments}



\begin{thebibliography}{150}

\bibitem{Aad:2012tfa}
  G.~Aad {\it et al.}  [ATLAS Collaboration],
  Phys.\ Lett.\ B {\bf 716}, 1 (2012).

\bibitem{Chatrchyan:2012ufa}
  S.~Chatrchyan {\it et al.}  [CMS Collaboration],
  Phys.\ Lett.\ B {\bf 716}, 30 (2012).




\bibitem{Dawson:2013bba}
  S.~Dawson, A.~Gritsan, H.~Logan, J.~Qian, C.~Tully, R.~Van Kooten, A.~Ajaib and A.~Anastassov {\it et al.},
  ``Working Group Report: Higgs Boson,''  arXiv:1310.8361 [hep-ex].



\bibitem{Lee:1977yc}
  B.~W.~Lee, C.~Quigg and H.~B.~Thacker,
  Phys.\ Rev.\ Lett.\  {\bf 38}, 883 (1977);
%
  Phys.\ Rev.\ D {\bf 16}, 1519 (1977).  


\bibitem{Cahn:1983vi}
  R.~N.~Cahn and M.~Suzuki,
  Phys.\ Lett.\ B {\bf 134}, 115 (1984).

\bibitem{Rupp:1991bb}
  G.~Rupp,
  Phys.\ Lett.\ B {\bf 288}, 99 (1992).  

\bibitem{Wurtz:2013ova}
  M.~Wurtz and R.~Lewis,
  Phys.\  Rev.\ D {\bf 88}, 054510 (2013).


\bibitem{Grinstein:2007iv}
  B.~Grinstein and M.~Trott,
  Phys.\ Rev.\ D {\bf 76}, 073002 (2007).

\bibitem{Barger:2012hv}
  V.~Barger, M.~Ishida and W.~-Y.~Keung,
  Phys.\ Rev.\ Lett.\  {\bf 108}, 261801 (2012).

\bibitem{Kaplan:1998tg}
  D.~B.~Kaplan, M.~J.~Savage and M.~B.~Wise,
  Phys.\ Lett.\ B {\bf 424}, 390 (1998);
  Nucl.\ Phys.\ B {\bf 534}, 329 (1998).

\bibitem{vanKolck:1998bw}
  U.~van Kolck,
  Nucl.\ Phys.\ A {\bf 645}, 273 (1999).

\bibitem{Jia:2004qt}
  Y.~Jia,
  hep-th/0401171.

\bibitem{Haidenbauer:2006dm}
  J.~Haidenbauer {\it et al.},
  Phys.\ Lett.\ B {\bf 643}, 29 (2006).


\bibitem{Dvoeglazov:1990bt}
  V.~V.~Dvoeglazov, V.~I.~Kikot and N.~B.~Skachkov,
  JINR-E2-90-569, 570.  
%

\bibitem{Gupta:1993tq}
  S.~N.~Gupta, J.~M.~Johnson and W.~W.~Repko,
  Phys.\ Rev.\ D {\bf 48}, 2083 (1993).

\bibitem{Sirlin:1980nh}
  A.~Sirlin,
  Phys.\ Rev.\ D {\bf 22}, 971 (1980).

\bibitem{Hollik:1988ii}
  W.~F.~L.~Hollik,
   Fortsch.\ Phys.\  {\bf 38}, 165 (1990).

\bibitem{Denner:1991kt}
  A.~Denner,
  Fortsch.\ Phys.\  {\bf 41}, 307 (1993).

\bibitem{Kublbeck:1990xc}
  J.~Kublbeck, M.~Bohm and A.~Denner,
  Comput.\ Phys.\ Commun.\  {\bf 60}, 165 (1990);  
  \\
  T.~Hahn,
  Comput.\ Phys.\ Commun.\  {\bf 140}, 418 (2001).

\bibitem{Mertig:1990an}
  R.~Mertig, M.~Bohm and A.~Denner,
  Comput.\ Phys.\ Commun.\  {\bf 64}, 345 (1991).  

\bibitem{Beneke:1997zp}
  M.~Beneke and V.~A.~Smirnov,
  Nucl.\ Phys.\ B {\bf 522}, 321 (1998).


\bibitem{Beringer:1900zz}
  J.~Beringer {\it et al.}  [Particle Data Group Collaboration],
  Phys.\ Rev.\ D {\bf 86}, 010001 (2012).  


\bibitem{Gasser:2009zz}
  J.~Gasser,
  PoS EFT {\bf 09}, 029 (2009).


\end{thebibliography}
\end{document}